\documentstyle[12pt,aaspp4]{article}
\received{~~} \accepted{~~}
\journalid{}{} \articleid{}{}

\begin{document}

\title{ASCA and contemporaneous ground-based observations \\
of the BL Lacertae objects 1749+096 and 2200+420 (BL Lac)}

\author{Rita M. Sambruna} 
\affil{The Pennsylvania State University, Department of Astronomy and
Astrophysics, 525 Davey Lab, State College, PA 16802}

\author{Gabriele Ghisellini}
\affil{Osservatorio Astronomico di Brera, via Bianchi 46, 22055 Merate
(LC), Italy}

\author{Eric Hooper} 
\affil{Harvard-Smithsonian Center for Astrophysics, 60 Garden St.,
Cambridge, MA 02138}

\author{R. I. Kollgaard} \affil{Fermi National Accelerator Laboratory,
Pine Road and Kirk Street, Batavia, IL 60510}

\author{Joseph E. Pesce} \affil{The Pennsylvania State University,
Department of Astronomy and Astrophysics, 525 Davey Lab, State
College, PA 16802}

\author{C. Megan Urry}
\affil{STScI, 3700 San Martin Dr., Baltimore 21218, MD}

\newpage 

\begin{abstract}

We present ASCA observations of the radio-selected BL Lacertae objects
1749+096 ($z$=0.32) and 2200+420 (BL Lac, $z$=0.069) performed in 1995
September and November, respectively. The ASCA spectra of both sources
can be described as a first approximation by a power law with photon
index $\Gamma \approx 2$. This is flatter than for most X-ray-selected
BL Lacs observed with ASCA, in agreement with the predictions of
current blazar unification models. While 1749+096 exhibits tentative
evidence for spectral flattening at low energies, a concave continuum
is detected for 2200+420: the steep low-energy component is consistent
the high-energy tail of the synchrotron emission responsible for the
longer wavelengths, while the harder tail at higher energies is the
onset of the Compton component.  The two BL Lacs were observed with
ground-based telescopes from radio to TeV energies contemporaneously
with ASCA.  The spectral energy distributions are consistent with
synchrotron-self Compton emission from a single homogeneous region 
shortward of the IR/optical wavelengths, with a second component in the
radio domain related to a more extended emission region. For 2200+420,
comparing the 1995 November state with the optical/GeV flare of 1997
July, we find that models requiring inverse Compton scattering of
external photons provide a viable mechanism for the production of the
highest (GeV) energies during the flare. In particular, an increase of
the external radiation density and of the power injected in the jet
can reproduce the flat $\gamma$-ray continuum observed in 1997 July.
A directly testable prediction of this model is that the line
luminosity in 2200+420 should vary shortly after ($\sim$ 1 month) a
non-thermal synchrotron flare.

\end{abstract}

\noindent {\underline{\em Subject Headings:}} Galaxies:active -- BL
Lacertae objects:individual (1749+096 and 2200+420) -- Blazars: X-ray
and $\gamma$-ray emission -- Blazars: multifrequency variability --
Radiation mechanisms: non-thermal.

\section{Introduction and Motivation}

Among Active Galactic Nuclei, blazars are the most dominated by
non-thermal activity from a relativistic jet oriented close to the
line of sight, with synchrotron emission from radio to UV/X-rays, and
inverse Compton up to $\gamma$-rays (e.g., Urry \& Padovani 1995;
Kollgaard 1994).  As such, they represent a fortuitous natural
laboratory to study the physics of the jets and, ultimately, the
mechanisms of energy extraction from the central black hole, a
fundamental goal of extragalactic astrophysics.

The blazar family is traditionally divided into two major subgroups,
depending on the strength of the optical emission lines: in Flat
Spectrum Radio Quasars (FSRQs) the lines are broad and strong
(rest-frame equivalent width EW$_r \gtrsim$ 5 \AA), while in BL
Lacertae objects (BL Lacs), where the lines are weak (EW $<$ 5 \AA) or
absent (e.g., Stickel et al. 1991).  However, the distinction into
FSRQs and BL Lacs based on the optical spectra has been recently
blurred by the detection of strong optical emission lines in several
sources traditionally classified as BL Lacs (Scarpa \& Falomo 1997),
including the class prototype, 2200+420 (BL Lac; Vermeulen et
al. 1995; Corbett et al. 1996), one of the targets discussed in this
paper.

Blazars in existing complete samples have subtly different spectral
energy distributions (SEDs) that give clues about the underlying jet
physics (Sambruna, Maraschi, \& Urry 1996).  The SEDs are
characterized by two broad spectral components, the first one
extending from radio to X-rays, and interpreted as due to synchrotron
radiation, and a second component from medium-hard X-rays to
$\gamma$-rays (up to TeV energies), attributed to inverse Compton
scattering of photons of various origins off the jet electrons (e.g.,
Mukherjee et al. 1997; Hartman 1996; von Montigny et al. 1995). A
practical way to parameterize the different SEDs is through the
radio-to-X-ray spectral index $\alpha_{rx}$, which is $> 0.8$ in
objects with the synchrotron peak in IR/optical (Low-energy peaked BL
Lacs, LBLs, and FSRQs) and $<$ 0.8 in UV/soft X-ray-peaked sources
(High-energy peaked BL Lacs, HBLs; Padovani \& Giommi 1995). The
synchrotron and inverse Compton components peak at different locations
in the SEDs depending on the bolometric jet luminosity, with a trend
of decreasing peak frequency (for both spectral components) with
increasing luminosity (Fossati et al. 1998; Sambruna et al. 1996).

The spectral transition from HBLs to LBLs to FSRQs can not be
explained in terms of different viewing angles alone, as postulated by
earlier unification schemes (e.g., Urry, Padovani, \& Stickel
1991). Instead, a change of jet physical parameters is required: the
higher synchrotron frequencies of HBLs suggest higher magnetic fields
and/or larger electron energies than in their LBL and FSRQ
counterparts (Sambruna et al. 1996; Georganopoulos \& Marscher
1998). In addition, the $\gamma$-ray loudness (i.e., the ratio of the
GeV to optical flux) increases from HBLs to LBLs to FSRQs, and is
correlated with the luminosity of the broad lines (Sambruna
1997). This suggests that a key role for the production of the higher
energies is played by the jet ambient radiation (Sambruna 1997), and
indeed this hypothesis is supported by a detailed modelling of the
spectral energy distributions of blazar types (Ghisellini et
al. 1998). However, strong selection effects, related to the
large-amplitude variability at $\gamma$-rays combined with the limited
EGRET sensitivity, could be at work at high energies exaggerating the
$\gamma$-ray dominance, particularly in the more luminous
quasars. More sensitive observations at GeV energies, e.g., with
GLAST, will provide a crucial test of the new unification scenario at
higher energies.

At lower energies, the ``modified'' blazar unification model is
directly testable through broad-band X-ray observations. As a
consequence of the synchrotron dominance, HBLs should exhibit steep
and downward-curved (as a result of radiative losses) X-ray spectra,
and indeed this was observed in ROSAT, EXOSAT, Ginga, and ASCA data
(Kubo et al. 1998; Sambruna et al. 1997, 1994a; Perlman et al. 1996;
Sembay et al. 1993). The observed 0.1--20 keV continua are typically
described by a power law with photon index $\Gamma \sim 2.4$,
steepening by $\Delta\Gamma \lesssim 0.5$ above a few keV.  Flux and
spectral variability are observed, with a trend of a harder continuum
with increasing intensity (e.g., Sambruna et al. 1994b), most likely
related to recent electron injection/acceleration events in the jet
(Ulrich et al. 1997).

In contrast, in LBLs and FSRQs flatter X-ray spectra are predicted as
a result of the larger Compton dominance. Soft X-ray observations with
ROSAT and {\it Einstein} confirm this prediction (Sambruna 1997;
Comastri et al. 1997; Urry et al. 1996; Worrall \& Wilkes 1990),
yielding typical slopes $\Gamma \sim 2$. Comparing the
(non-simultaneous) ROSAT (0.1-2.4 keV) and {\it Einstein} IPC
(0.3--3.5 keV) data for 13 LBLs and FSRQs, spectral upturns were
inferred above $\sim$ 1 keV in both subclasses (Sambruna 1997; Urry et
al. 1996), and directly observed only in one object so far (0716+714;
Makino et al. 1996). In a few LBLs the observed correlated flux-slope
variability trend is in the sense of {\it flatter} soft X-ray spectra
for {\it decreasing} flux, as if when the synchrotron intensity fades
the flatter Compton tail is uncovered.  For FSRQs, little or no flux
and spectral variations (20\% or less) are found from repeated ROSAT
observations on timescales of days to months (Sambruna 1997),
consistent with Compton dominance at X-rays in these blazars.

At hard X-rays, contrary to HBLs information has been scarce so far
for LBLs. Five objects observed with EXOSAT at energies $\gtrsim 2$
keV showed rather flat spectra, $\Gamma \sim 1.5$ (Sambruna et
al. 1994a,b), albeit within large errors. Ginga observations of a few
sources also indicated rather flat slopes, but with a large dispersion
(Ohashi et al. 1989). More observations are clearly needed to assess
the continuum emission at medium-hard X-rays of LBLs and probe the
modified unification model.

To this purpose, we were granted ASCA time during AO3 to observe the
two sources 1749+096 ($z=0.320$) and 2200+420=BL Lac ($z=0.069$). Both
sources belong to the complete 1 Jy sample of BL Lacs assembled at
radio frequencies (Stickel et al. 1991), and can be classified as LBLs
on the basis of their $\alpha_{rx}$ indices ($\alpha_{rx}=0.92$ for
1749+096 and $\alpha_{rx}$=0.85 for 2200+420; Sambruna et al. 1996).
They were selected because they are the X-ray-brightest LBLs with
little or no spectral information at hard X-rays. In the case of
2200+420, the comparison of the ROSAT and IPC spectra indicated an
upturn of the continuum at $\gtrsim 1$ keV (Urry et al. 1996),
suggesting that we might detect the Compton component with ASCA.

Another goal of our investigation is the detailed study of continuum
emission processes. To this end we organized ground-base coverage at
radio through optical frequencies simultaneous to ASCA for both
sources. By a coincidence, 2200+420 was also observed at TeV energies
with Whipple during our ASCA pointing, yielding only an upper limit on
the TeV flux.

Recently, 2200+420 came into much attention since it underwent a large
outburst at GeV and optical frequencies during 1997 July (Bloom et
al. 1998). This LBL is also remarkable since a strong (L$_{H\alpha}
\sim 10^{41}$ erg s$^{-1}$) and broad (FWHM $\sim$ 4,000 km s$^{-1}$)
H$\alpha$ emission line was detected in 1995 May-June (Corbett et
al. 1996; Vermeulen et al. 1995), which appears to be variable on
short ($\sim$ 1 week) timescales (Robinson 1996; Nesci \& Massaro
1996). Interestingly, the line detection occurred a few months after
the first EGRET detection of the source (Catanese et al. 1997). In
contrast, 1749+096 was never detected in $\gamma$-rays and shows weak
lines (Stickel et al. 1991).

This paper is organized as follows. The ASCA data are described in
\S~2, together with the results of the time and spectral analysis. We
find that the 0.6--10 keV continua of both LBLs are flat ($\Gamma \sim
2.0$), with further spectral complexity over a single power law
model. In the case of 2200+420, we detect a concave continuum with a
hard tail emerging above 2 keV, while 1749+096 shows low-energy
spectral flattening.  The contemporaneous ground-based observations
are presented in \S~3. We construct multifrequency spectral energy
distributions and compare them to published spectra (including the
1997 GeV/optical flare for 2200+420). Our results are summarized and
discussed in \S\S~4 and 5, with the conclusions following in
\S~6. Throughout this paper, H$_0$=50 km s$^{-1}$ Mpc$^{-1}$ and
q$_0$=0.5 are assumed. The energy index $\alpha$ is defined as
F$_{\nu} \propto \nu^{-\alpha}$. 

\section{ASCA Observations} 

\subsection{Data Reduction and Analysis}

We observed 1749+096 and 2200+420 with ASCA on 1995 September 22 and
on 1995 November 22, respectively, for 30 ksec each.  For a
description of the ASCA experiment see Tanaka, Inoue, \& Holt (1994).
In both cases, the Solid-State Imaging Spectrometers (SIS0 and SIS1)
operated in 1-CCD \verb+FAINT+ mode and the Gas Imaging Spectrometers
(GIS2 and GIS3) were used in Pulse-Height mode.  In order to apply
standard data analysis methods, the \verb+FAINT+ SIS mode was
converted into \verb+BRIGHT2+, applying the corrections for echo
effect and dark frame error (see ``The ASCA Data Reduction Guide'',
v.2, April 1997). 

The data reduction was performed using \verb+FTOOLS+ v.4.0.  The
screening criteria for both SIS and GIS included the rejection of the
data taken during the passage of the South Atlantic Anomaly and for
geomagnetic cutoff rigidity lower than 6 GeV/c.  We retained SIS data
accumulated for Bright Earth angles $>$ 20$^{\circ}$ and Elevation
angles $> 10^{\circ}$, and GIS data accumulated for Elevation angles
$> 5^{\circ}$. Only data corresponding to SIS grades 0, 2, 3, and 4
were accepted.  Table 1 reports the effective exposures after the
screening and the corresponding source count rates in the four
detectors.

The source spectra and light curves were extracted from circular
regions centered on the source position with radii of 4 arcmin for the
SIS and 6 arcmin for the GIS, which has a larger intrinsic point
spread function. In the case of the GIS we evaluated the background in
circular regions of radius 4 arcmin, located about 12 arcmin away from
the target where the source counts are negligible. In the case of the
SIS, the background was evaluated in circular regions of radius 2
arcmin located 6 arcmin away from the target on the same chip. Since
no spectral variability is apparent within the ASCA exposures, the
spectra were integrated over the entire duration of the observations.

The ASCA spectra were fitted using \verb+XSPEC+ v.10. The SIS and GIS
spectra were rebinned in order to have a minimum of 20 counts in each
spectral bin to validate the use of the $\chi^2$ statistic. In order
to increase the signal-to-noise ratio we performed joint fits to the
data from the 4 detectors, leaving only the normalizations as
independent parameters. The 1994 May response matrices were used for
the GIS spectra, while for the SIS data we used the matrices generated
by the \verb+SISRMG+ program (v.1.1, 1997 March). 
The SIS and GIS data were fitted in the energy ranges 0.6--10 keV and
0.7--10 keV, where the spectral responses are best known.

\subsection{Timing Analysis}

The ASCA light curves of both objects were inspected for flux
variability. For 1749+096 no significant flux variations were observed
in either the SIS nor the GIS light curves. The $\chi^2$ test yields a
probability that the intensity is constant at P$_{\chi^2} \gtrsim$
36\% and $\gtrsim$ 10\% for the SIS and GIS data, respectively, over a
timescale of a few ($\lesssim$ 7) hours.

In the case of 2200+420, flux changes were observed in the SIS0, SIS1,
and GIS2 light curves, with P$_{\chi^2} \sim$ 1.2\%, 0.4\%, and 2\%,
respectively, while no variability is detected in the GIS3 data
(P$_{\chi^2} \sim$ 62\%). Inspection of the light curves shows that,
while the background is stable for SIS0, SIS1, and GIS3, it is
variable in the case of GIS2, with P$_{\chi^2} \sim$ 20\%. The largest
flux variability in the light curves of SIS0, SIS1, and GIS3 is
present at the beginning of the pointing: the flux increases by 25\%
in the first $\sim$ 3 hours of the observation. We regard this result
with caution, since instabilities in the pointing manoeuver of the
satellite could introduce spurious flux changes. In addition,
P$_{\chi^2}$ appears to depend on the light curve binning, with larger
probabilities for smaller binsizes (and thus lower signal-to-noise
ratios). Our conclusion is that there is only tentative evidence for
significant flux variability in the ASCA light curves of 2200+420.

\subsection{Results of the Fits to the ASCA Spectra} 

The ASCA data were first fitted with a single power law absorbed at
the low energies by a column density of cold gas N$_H$ using the
Morrison \& McCammon (1983) cross section for photoelectric
absorption, with the abundances of neutral elements other than
hydrogen fixed at solar values.  In the spectral fits N$_H$ was left
both free and fixed to the Galactic column density along the line of
sight (see below).  The results of the spectral fits are reported in
Table 2 for both sources; the quoted uncertainties are the 90\%
confidence errors for one parameter of interest 
($\Delta\chi^2$=2.7). All quantities in Table 2 are in the observer's
frame.

Figures 1 and 2 show the ASCA spectra of 1749+096 and 2200+420,
together with the folded power law + Galactic N$_H$ model (panel a),
and the residuals in the form of the ratio of the data to the model
(panel b). Figure 3 shows the confidence contours for the power law
photon index, $\Gamma$, and the fitted column density N$_H$. The
vertical lines mark the Galactic N$_H$ and its nominal 90\%
uncertainty range.

Fits with more complex models were also attempted, as required by the
data. The significance of the improvement of the fit when additional
free parameters are included was evaluated using the F-test, assuming
as a threshold probability for a significant improvement P$_F$=95\%.

\subsubsection{1749+096} 

The Galactic column density in the direction to 1749+096 is N$_H^{Gal}
= (9.6 \pm 1.0) \times 10^{20}$ cm$^{-2}$ (Stark et al. 1992), where
the uncertainty corresponds to the 90\% confidence level (Elvis et
al. 1989). The ASCA data are well fitted by a single power law with
N$_H$ fixed to the Galactic value (Table 2); the fitted slope is
rather flat, $\Gamma \sim 1.8$, i.e., rising in energy emitted per
decade ($\nu$F$_{\nu}$). Figure 1 shows the ASCA data with the
best-fit folded model (top) and the ratio of the data to the model
(bottom). 

However, a better fit is obtained when the column density is allowed
to vary ($\Delta\chi^2$=13.2 for one additional free parameter). The
fitted column density, N$_H = 1.6^{+0.40}_{-0.34} \times 10^{21}$
cm$^{-2}$, is larger than Galactic at $\sim$ 98\% confidence (Fig. 3),
indicating excess N$_H$ along the line of sight. While excess N$_H$
could be at least in part due to a systematic effect of the SIS
(Dotani et al. 1996), two short ROSAT PSPC observations of the source
also indicated excess column densities, albeit within large
uncertainties (Urry et al. 1996). Alternatively, the spectral
flattening at soft energies could be due to either a convex continuum
or an absorption feature.

Fitting the ASCA data with a power law plus Galactic N$_H$ plus an
absorption edge gives a formally good fit, with reduced $\chi^2$ of
$\chi^2_r$=0.87 for 615 degrees of freedom, highly improved over the
single power law ($\Delta\chi^2$=20 for two additional parameters,
significant at P$_F \gtrsim$ 99\%). The fitted edge energy and optical
depth are E=0.73$^{+0.09}_{-0.53}$ keV and
$\tau=0.42^{+0.28}_{-0.22}$, respectively.  The edge best-fit energy
is consistent with the L-edge of FeIV(0.943 keV) or FeV(0.973 keV) in
the rest-frame of the source (assuming $z$=0.32), or, alternatively,
with the K-edge of OVIII (0.871 keV) from an intervening absorber at
$z=0.19$. However, the edge energy is poorly determined and its 90\%
lower limit is well below the ASCA sensitivity range. This more likely
suggests that the edge is just a parameterization of low-energy
spectral flattening due, for example, to an intrinsically curved
continuum.  Fitting the ASCA data with a broken power law model with
fixed Galactic column density yields a highly improved (P$_F \gtrsim$
99\%) fit with respect to the single power law with both fixed and
free N$_H$. A convex continuum, with steepening $\Delta\Gamma \equiv
\Gamma_2 - \Gamma_1 \sim$ 0.4 above E$_0$=2.2 keV, is indicated (Table
2).

Our ASCA data represent the first observations of 1749+096 at
medium-hard X-rays. Previously, the source was observed at soft
energies with the {\it Einstein} IPC (Worrall \& Wilkes 1990) and with
the ROSAT PSPC (Urry et al. 1996) during an intensity state lower than
ASCA in 0.5--2 keV. In both cases, the spectral parameters are poorly
constrained and are consistent with ASCA within the large
uncertainties.

In summary, the ASCA spectrum of 1749+096 is consistent with a hard
($\Gamma \sim 1.8$) power law. Tentative evidence for low-energy
spectral flattening, which can be described either with a single power
law with $\Gamma \sim 1.9$ and excess cold absorption over Galactic by
$\Delta$N$_H \sim 7 \times 10^{20}$ cm$^{-2}$, or by a convex broken
power law with steepening $\Delta\Gamma \sim 0.4$ above 2 keV.

\subsubsection{2200+420} 

Since 2200+420 is at low Galactic latitudes (b$^{II}$=--10.43), the
Galactic column density of neutral hydrogen from 21 cm measurements in
its direction is quite high, ($2.02 \pm 0.01) \times 10^{21}$
cm$^{-2}$ (Elvis et al. 1989).  In addition, the source is located
behind a molecular cloud from which CO emission and absorption has
been detected (Bania, Marscher, \& Barvainis 1991; Marscher, Bania, \&
Wang 1991; Lucas \& Liszt 1993). From recent estimates, the equivalent
atomic hydrogen column density of the CO cloud is N$_H \approx 1.6
\times 10^{21}$ cm$^{-2}$ (Lucas \& Liszt 1993). Thus, the total
Galactic column density in the direction to 2200+420 (atomic +
molecular) is $\approx 3.6 \times 10^{21}$ cm$^{-2}$. 

As apparent from Table 2, the fit with fixed Galactic absorption is
acceptable, $\chi^2_r$=0.97/798. Inspection of the residuals in Figure
2 shows, however, the presence of excess flux below $\sim$ 1 keV,
suggesting a concave spectrum.  Leaving the N$_H$ free to vary yields
a significantly better fit (P$_F \gtrsim$ 99\%), with column densities
in the range 2.3--3.0 $\times 10^{21}$ cm$^{-2}$ (Fig. 3).  The fitted
N$_H$ is lower than the total Galactic column density, suggesting a
concave continuum. Alternatively, if the ASCA-fitted column density
represents the total true value of the Galactic N$_H$ along the line
of sight, a contribution from the molecular cloud of N$_H \sim 7
\times 10^{20}$ cm$^{-2}$ is inferred, lower than current estimates
(Lucas \& Liszt 1993). 

We next fitted the 0.6--10 keV data with a broken power law with total
Galactic N$_H$ fixed to $3.6 \times 10^{21}$ cm$^{-2}$. The fit with
this model is highly improved ($\Delta\chi^2$=45 for two additional
parameters, P$_F >$ 99\%) with respect to the fit with the single
power law and Galactic column density, and even with respect to the
power law with free N$_H$ ($\Delta\chi^2$=9 for 1 additional
parameter, P$_F \gtrsim$ 99\%). The fit yields a low-energy photon
index $\Gamma_1 \sim 2.3$ and a harder slope $\Gamma_2 \sim 2$ above
$\sim$ 2 keV (Table 2). We interpret the steep component at $\lesssim$
2 keV as the high-energy end of the synchrotron emission responsible
for the longer wavelength emission (\S~3.2.2), while the flatter tail
at higher X-ray energies is the onset of the Compton component
extending to $\gamma$-rays. 

Our ASCA data confirm for the first time the previous inference, based
on non-simultaneous ROSAT and IPC data, that the X-ray continuum of
2200+420 has an upturn at $\sim$ a few keV (Urry et al. 1996).
2200+420 was also observed at energies $\gtrsim$ 1 keV with the {\it
Einstein} MPC and Ginga (Bregman et al. 1990; Kawai et al. 1991), when
the flux was slightly lower than ASCA (factor 1.7).  The MPC slope is
consistent with ASCA, while the two Ginga observations, separated by
one month, yield a harder ($\Gamma \sim 1.7$) and a steeper ($\Gamma
\sim$ 2.2) continuum, respectively. A discussion of correlated
flux-spectral variability is hampered by the different energy ranges
of the various satellites; in general, a variability trend of
flatter-when-fainter is apparent. However, observations with RXTE and
ASCA during the flare of 1997 July, when the flux was a factor 3
higher than in 1995 November, show a flatter ($\Gamma \sim 1.4$)
continuum up to 200 keV (Madejski, Jaffe, \& Sikora 1997; Makino et
al. 1997), contrary to this trend.

There is mounting evidence for the presence of intrinsic absorption in
the X-ray spectra of BL Lacs, at energies consistent with absorption
by ionized oxygen and with column densities as low at $\tau \sim 0.2$
(Sambruna \& Mushotzky 1998; Sambruna et al. 1997). We checked the
ASCA data of 2200+420, which has a low redshift ($z$=0.069), for the
presence of ionized oxygen absorption by adding an edge with an energy
constrained to vary in the range of OVI to OVIII, i.e., 0.63--0.81 keV
in the observer's frame. No statistically significant improvement was
obtained, with a formal upper limit to the optical depth of the edge
of oxygen (in any ionization state) of $\tau < 1.3$ at 90\%
confidence.

In summary, the ASCA observations of 2200+420 in 1995 November are
best-fitted by a concave broken power law with a steep photon index
($\Gamma \sim 2.3$) below 2 keV, attributed to the synchrotron
emission, and a flatter tail ($\Gamma \sim 2$) at harder energies,
related to the inverse Compton component. This is the first time that
both spectral components are detected in the X-ray spectrum of
2200+420.

\section{Observations Contemporaneous to ASCA}

\subsection{Radio}

The Very Large Array was used to observe 1749+096 and 2200+420 in late
1995, with core radio fluxes being measured at five frequencies during
each session.  Table 3A reports the log of the radio observations.
The observations of 1749+096 were made while the VLA was in transition
from the A to B-configurations, and all of the 2200+420 data were
taken in the latter configuration.  The data were analyzed in the
standard manner using the NRAO's Astronomical Image Processing System
(\verb+AIPS+).  The initial calibration was made using either 3C286 or
3C48 as the primary flux standard.  Data were recorded in two 50 MHz
bandpasses at each frequency, and analyzed simultaneously using the
\verb+AIPS+ tasks \verb+IMAGR+ and \verb+CALIB+ for deconvolution and
self-calibration, respectively.  Two iterative cycles of \verb+IMAGR+
and \verb+CALIB+ were used for each data set, with corrections applied
only to the source phases in order to preserve the absolute flux
density scale.

The resulting images show no extended emission around the unresolved
cores of either source, and therefore the measured flux densities were
not affected by the changing resolution of the VLA. The rms error on
each image was less than 1\%, and the errors in the flux density are
dominated by uncertanties in the overall flux calibration.  We
estimate these errors to be a few percent at the lower frequencies,
and $\sim$ 5\% at 22 GHz.  A larger uncertainty is expected from the
two observations that used 3C48 as the primary calibrator (1749+096 on
1995 September 28 and 1995 December 5) because of difficulties in
obtaining an adequate calibrator solution.  The 1995 November 28
observations of 2200+420 were marred by corrupted data which rendered
all primary calibrator sources unuseable.  Our estimates of the flux
density for these data are based on the raw, uncalibrated data which
should have a flux scale approximately 10\% of the true scale.  We
have therefore increased our error estimate for the flux densities
measured under these circumstances.  The radio flux densities are
reported in Table 3A.

\subsection{Optical}

\subsubsection{Observations and Data analysis}

Both BL Lac objects were observed at optical wavelengths in the Fall
of 1995 in conjunction with the ASCA observations.  CCD images in $V$
and $R$ were obtained for 1749+096 on the night of UT September 23,
and two months later 2200+420 was imaged in $BVRI$ on UT November 23.
All data were collected with the $61''$ reflector on Mt. Bigelow,
Arizona.  Only three measurements were made of 1749+096, but 2200+420
was monitored over a span of five hours. No variability was detected,
and so the magnitudes were averaged over each night. Aperture
photometry and growth curve analysis were performed with Stetson's
(1987, 1990) \verb+DAOPHOT+ and \verb+DAOGROW+ programs.

Stray light contaminated the images on both nights, introducing
spatially and temporally variable additive features.  The additive
nature of this light was confirmed by examining the observed
magnitudes of calibrated stars in globular clusters (Christian et
al. 1985; Odewahn et al. 1992) which filled the field of view of the
CCD camera.  Each BL Lac field contains several stars that have been
calibrated to facilitate the monitoring of the AGNs (Craine, Johnson,
\& Tapia 1975; Smith et al. 1985; Fiorucci \& Tosti 1996).  In the
present case, these calibration stars were used to check the validity
of the data reduction and to estimate the uncertainties of the derived
magnitudes.  Both nights were photometric, so the external globular
cluster calibrators were used to establish the transformation from
instrumental to standard magnitudes.  The derived magnitudes of the
calibrated field stars were consistent with the values in the
literature for 1749+096 and about 0.05 mag fainter for 2200+420 .
(The calibrated values for the latter object have systematic
differences of up to 0.1 mag; see Fiorucci \& Tosti 1996.)  The flux
values of 2200+420 were made brighter by 0.05 mag to compensate.  Both
sets of calibration stars have an rms scatter of $\approx 0.1$ mag
about the values given in the literature; 0.1 mag was taken to be the
uncertainty in the flux measurements of the program objects.

The observed magnitudes are listed in Table 3B. The second entry in
the Table lists the corresponding intrinsic flux densities, obtained
after dereddening the magnitudes with A$_V$=0.7 for 1749+096 and
A$_V$=1.2 for 2200+420. The extinction values were calculated
following the law of Cardelli, Clayton, \& Mathis (1989), assuming
R$_V$=3.1 (the latter value, however, is uncertain and could be as
high as 5). For 1749+096, A$_V$ was derived from the ASCA-fitted N$_H$
(in the conservative assumption that spectral bending is due to excess
absorption) and using the conversion A$_V=4.5 \times 10^{-22}$N$_H$
mag cm$^2$ (Ryter 1996); the derived value, A$_V$=0.7, is 43\% higher
than derived from Galactic N$_H$ only, A$_V$=0.4.  For 2200+420, a
similar calculation using total Galactic N$_H=3.6 \times 10^{21}$
cm$^{-2}$ gives A$_V=1.6$; fitting an optical spectrum of the source,
Vermeulen et al. (1995) find that a lower value A$_V$=0.9 provides
excellent results, with {\it a posteriori} consistency arguments.  We
used A$_V=1.2$ as an intermediate value. The flux density errors
listed in Table 3B include only the photometric imprecisions, with no
contributions from ill-determined uncertainties in the extinction
parameters.  The effects of variations in the extinction values were
explored in calculations of optical spectral indices.

\subsubsection{Optical indices and $\alpha_{ox}$ indices} 

We calculated optical energy indices $\alpha_{opt}$ from a linear fit
to the dereddened flux densities, in order to compare $\alpha_{opt}$
with the ASCA slope and gain insights about the origin of the X-rays.
For 2200+420, a spectral index of $\alpha_{opt} = 0.76 +/- 0.20$ was
derived assuming the fluxes in Table 3B which were dereddened using
A$_V=1.2$ and R$_V$=3.1 (see above). This value of $\alpha_{opt}$ is
slightly flatter than the ASCA slope measured between 0.6--2 keV,
$\alpha_{X,soft}=1.31 \pm 0.08$ (Table 2), indicating that further
steepening could be occurring between the optical and the X-ray bands,
in agreement with synchrotron emission where radiative losses are
responsible for depleting the emission at the higher energies first.
However, the optical slope is sensitive to the adopted reddening.
Increasing $R_V$ to the rather high value of 5 or lowering $A_V$ to
0.9 both produced $\alpha_{opt} \sim 1.1$, consistent with the ASCA
slope, while higher extinction, $A_V = 1.6$, flattened the slope to
$\alpha_{opt} \sim 0.31$, again flatter than the X-ray slope.

For 1749$+$096, we obtain $\alpha_{opt} = 0.08 +/- 1.2$ from the two
optical measurements in Table 3B, very poorly constrained due to the
large uncertainty of the $V$ magnitude and the small wavelength
difference between the $V$ and $R$ bands.  Altering $A_V$ and $R_V$
had negligible effect compared to the error on $\alpha_{opt}$.

The composite spectral index, $\alpha_{ox}$, connecting the optical
and X-rays (defined between 5.5 $\times 10^{14}$ Hz and 1 keV), turns
out to be less dependent on the reddening and thus a more reliable
indicator of the synchrotron slope above the optical peak. For
2200+420, using the optical $V$ flux in Table 3B and the 1 keV X-ray
flux in Table 2 we derive $\alpha_{ox}=1.29 \pm 0.02$, in excellent
agreement with $\alpha_{X,soft}$. Using the optical fluxes dereddened
with A$_V$=1.6 and A$_V=0.9$ we obtain $\alpha_{ox}=1.24 \pm 0.02$ and
$\alpha_{ox}=1.35 \pm 0.02$, consistent within the errors with
$\alpha_{X,soft}$. We thus conclude that the similarity of the
optical-to-X-ray and soft X-ray (0.5--2 keV) slopes provides support
for our earlier claim that the steep power law component detected in
the ASCA data of 2200+420 is the high-energy tail of the synchrotron
component responsible for the lower energies.

For 1749+096, we derive $\alpha_{ox} = 1.36 \pm 0.05$, steeper than
the ASCA slope, using the optical flux in Table 3B. Using the optical
flux dereddened with a lower A$_V=0.4$, we obtain $\alpha_{ox}=1.32
\pm 0.05$, again steeper than ASCA. This indicates that the X-rays
belong to a different spectral component than the lower frequencies,
most likely the onset of a Compton scattering component.

\subsection{$\gamma$-rays}

By a fortunate coincidence, 2200+420 was observed at TeV energies with
the Whipple Observatory between 1995 October 17 and November 25,
overlapping with our ASCA pointing.  Only an upper limit of $5.3 \times
10^{-10}$ photons cm$^{-2}$ s$^{-1}$ for to the flux above 350 GeV was
obtained (Catanese et al. 1997). 
At GeV energies, 2200+420 was first detected with EGRET in 1995
January--February, $\sim$ 9 months before the ASCA and Whipple
observations, with a significance of 4.4$\sigma$ (Catanese et
al. 1997). A large flare at GeV energies was later detected in 1997
July (Bloom et al. 1998); the flux was a factor $\sim$ 4 higher than
the 1995, with a flare of amplitude 60\% in $\sim$ 8 hrs. The EGRET
spectrum was significantly flatter during the flare, $\Gamma=1.68 \pm
0.12$, than during the lower level of early 1995, when $\Gamma=2.27
\pm 0.30$ (Bloom et al. 1998). This confirms the trend of harder
$\gamma$-ray spectra with increasing luminosity which has also been
observed in other $\gamma$-ray blazars (Mukherjee et al. 1997).

\subsection{Multifrequency Spectral Energy Distributions}

The radio to $\gamma$-ray spectral energy distributions (SEDs) of
1749+096 and 2200+420 from our contemporaneous ASCA and ground-based
observations are shown in Figures 4 and 5, respectively.  For
1749+096, which was never detected at $\gamma$-rays, the plotted upper
limit to the GeV flux is equal to the EGRET sensitivity
threshold. Plotted with different symbols are also SEDs derived from
published data for both sources.  For 1749+096, we plotted
non-simultaneous literature data, representing an ``average'' spectral
distribution. For 2200+420, the SEDs from the 1988 campaign (Bregman
et al. 1990; Kawai et al. 1991) and from the 1997 GeV/optical outburst
(Bloom et al. 1998; Madejski et al. 1997; Makino et al. 1997) are
shown. The optical datapoints were dereddened using A$_V$=0.7 for
1749+096 and A$_V$=1.2 for 2200+420 (\S~3.2.1).

As expected (see \S~1), the radio to X-ray flux distributions of both
sources peak at low frequencies, below the X-ray regime. Following
Sambruna et al. (1996), we performed a parabolic fit to the radio
through optical/X-ray 1995 SEDs of both sources (filled symbols) to
determine the value of the peak frequency, $\nu_p$, as discussed in
detail in Appendix 1. We obtain $\nu_p=1.4 \times 10^{13}$ Hz for
1749+096 and $\nu_p=2.2 \times 10^{14}$ Hz for 2200+420, comparable to
the values derived by Sambruna et al. (1996) based on literature
data. The same fits yield an estimate of the luminosity at the peak
frequency, L$_p$, and of the integrated radio-to-X-ray synchrotron
luminosity, L$_{sync}$: for 1749+096, L$_p=1.6 \times 10^{33}$ erg
s$^{-1}$ Hz$^{-1}$ and L$_{sync}=1.1 \times 10^{47}$ erg s$^{-1}$, for
2200+420 L$_p=4 \times 10^{29}$ erg s$^{-1}$ Hz$^{-1}$ and L$_{sync}=3
\times 10^{44}$ erg s$^{-1}$. Uncertainties are 35\% for L$_{sync}$,
30\% for L$_p$, and 10\% for $\nu_p$ (Appendix 1).

In the case of 2200+420, comparison of the ``quiescent'' 1995 state
with the 1997 flare shows that large flux and spectral variability
occurred above a few keV. The X-ray flux increased by a factor 3 and
hardened from $\Gamma \sim 2.0$ to $\Gamma \sim 1.4$. A similar
behavior of hardening of the spectrum with increasing intensity is
also observed in the EGRET band when the 1995 January and the 1997
July EGRET spectra are compared. Moreover, the flatter 1997
$\gamma$-ray spectrum suggests a second peak in the Compton component
at $\gtrsim$ 10$^{24}$ Hz in addition to the peak around 10$^{22}$ Hz.

In the case of 1749+096, the 1995 SED corresponds to a high state in
optical and X-rays with respect to historical spectra. The ASCA data
lie above the extrapolation of the lower frequencies (\S~3.2.2),
indicating that a second spectral component must be present in the
SED, although at a level currently under the EGRET sensitivity limit
(arrow in Figure 4). We will return to the spectral distributions of
the two sources in \S~5.2.

\section{Summary of the Observational Results}

We have presented ASCA and contemporaneous radio through $\gamma$-ray
observations of the two LBLs 1749+096 and 2200+420 (=BL Lac). Our
principal observational results are summarized as follows:

\begin{itemize}

\item The ASCA spectra of 1749+096 and 2200+420 can be described by 
single power laws with photon indices $\Gamma \sim 1.8$ and $\Gamma
\sim 2.1$, respectively.

\item 1749+096 exhibits tentative evidence for spectral flattening at
low energies. A convex broken power law, with steepening
$\Delta\Gamma=0.4$ above 2.2 keV, provides a good fit to the data.
Alternatively, there is excess absorption by cold gas ($\Delta$N$_H
\sim 7 \times 10^{20}$ cm$^{-2}$) along the line of sight.

\item There is evidence for a concave spectrum in 2200+420, with a
steep component ($\Gamma \sim 2.3$) below $\sim$ 2 keV and a harder
tail ($\Gamma \sim 2$) at higher energies. 

\item Contemporaneous radio to optical measurements confirm that the
spectral energy distribution of both sources peak at $\sim 10^{13-14}$
Hz, as expected since they are LBLs. 

\item During the 1997 July GeV/optical flare, the X-ray continuum of
2200+420 flattened dramatically with respect to our ``quiescent''
state of 1995 November. The $\gamma$-ray spectrum was flatter than
during 1995 January. 

\end{itemize}

\section{Discussion}

\subsection{ASCA Spectra of Blazars: Implications for the 
Unification Models}

In its operational lifetime, ASCA has observed up to now several
blazars of the HBL, LBL, and FSRQ types. A subset of objects observed
simultaneously with EGRET, including 7 HBLs, 4 LBLs, and 7 FSRQs, is
presented in Kubo et al. (1998).  We add the two LBLs studied here,
1749+096 and 2200+420, and three FSRQs with published ASCA data,
0836+71 and 0438--436 (Cappi et al. 1997), and PKS 1510--089 (Singh,
Shrader, \& George 1997).

The ASCA spectra of the FSRQ subgroup are generally well described by
a single power law model; the fitted photon indices span the range
$\Gamma=1.3-1.8$ and have mean $\langle \Gamma_{FSRQ} \rangle = 1.59$
and dispersion $\sigma_{FSRQ}$=0.14. The ASCA spectra of HBLs are to a
first approximation described by a single power law with slopes in the
range $\Gamma=2.0-2.7$, with mean $\langle \Gamma_{HBL} \rangle =
2.47$ and dispersion $\sigma_{HBL}=0.36$. In contrast, for LBLs
$\Gamma$ is in the range 1.6--2.1, with $\langle \Gamma_{LBL} \rangle
= 1.85$ and $\sigma_{LBL}=0.16$.  Our tentative conclusion, based on
limited statistics, is that ASCA observations of blazars confirm the
expectations of the modified unification models (\S~1): while HBLs
have in general steep hard X-ray continua, dominated by the
synchrotron high-energy tail, LBLs and FSRQs are characterized by
flatter slopes, as expected if the inverse Compton component becomes
dominant (but see below for FSRQs). Clearly, in order to draw
definitive conclusions larger, possibly complete, samples are needed,
especially of LBLs, which we anticipate from the public archives
within the next few years.

Indeed, in the LBLs 2200+420 and 0716+714, which are both GeV sources,
the synchrotron and inverse Compton component are explicitly resolved
at X-rays: in both cases the ASCA spectra are best-fitted by a concave
broken power law, with a steep power law at softer energies and a flat
tail emerging above $\sim$ 2 keV (Table 2 and Makino et al. 1996,
respectively). Similarly, ROSAT observations contemporaneous to ASCA
for another GeV-bright LBL, 0235+164, are consistent with a steeper
component in a range of softer energies (Madejski et al. 1996). It is
interesting that the $\alpha_{rx}$ indices of these three LBLs are in
the range 0.75--0.85, i.e., close to the dividing line between HBLs
and LBLs; they could thus be ``transitional'' cases, which would
explain their concave X-ray spectra as a mix of synchrotron and
inverse Compton.

The ASCA data for FSRQs are consistent with flat X-ray continua, at
the hard end of the distribution for the blazar class. However, it is
important to realize that the FSRQs so far observed with ASCA are
either at large redshifts, $z > 1$, and/or are strong GeV $\gamma$-ray
emitters (e.g., Thompson et al. 1995). Given the anticorrelation
between the X-ray slope and $z$ (e.g., Sambruna 1997), and the fact
that GeV blazars are dominated at higher energies by a hard Compton
tail, there is a clear bias for flat X-ray slopes for the current ASCA
sample of FSRQs. Thus the results for FSRQs are still inconclusive for
what concerns blazar unified models; a clearer test will come from
medium-hard X-ray observations of low-$z$ (non-GeV) FSRQs. Indeed, our
study of the ROSAT pointed spectra of a large (41 objects) sample of
FSRQs (with $z$=0--3) already shows that these sources have a wide
distribution of soft X-ray slopes, $0.5 \lesssim \Gamma \lesssim 2.5$,
indistinguishable from LBLs and with a large overlap with HBLs at
matching redshifts (Sambruna 1997; see also Padovani, Giommi, \& Fiore
1997).  This suggests the existence of a subclass of steep soft X-ray
FSRQs with possibly unusual intrinsic conditions. Moreover, our recent
deep X-ray survey of blazars (Wide Angle ROSAT Pointed Survey; Perlman
et al. 1998) is finding more and more examples of FSRQs with HBL-like
spectral distributions; this new FSRQ subclass represents a sizable
fraction (25\%) of the total FSRQ population. It will be important to
study with ASCA and SAX their hard ($\gtrsim$ 2 keV) X-ray continua to
verify whether their high-energy continua are steep and convex, as for
HBLs, or whether a flat component arises, similar to LBLs and high-$z$
FSRQs, and to determine the true origin of the soft X-ray flux
(non-thermal synchrotron from the jet versus thermal emission from the
disk). Multifrequency observations of these ``HBL-like'' FSRQs are
also needed to study in detail the continuum emission processes and
clarify the role of this new blazar class for the unified models.

\subsection{Interpretation of the SEDs of 1749+096 and 2200+420} 

\subsubsection{General Considerations} 

While there is consensus that the smooth, variable, and polarized
radio-to-UV emission from blazars is due to synchrotron emission from
high-energy electrons in relativistic motion (e.g., Angel \& Stockman 
1980; Blandford \& Rees 1978), the detailed mechanisms for the
production of the high-energy continuum are still unclear.  The X-ray
to $\gamma$-ray continuum is generally attributed to inverse Compton
(IC) scattering of either the synchrotron photons themselves
(synchrotron-self Compton, SSC; Maraschi, Ghisellini, \& Celotti 1992)
or ambient thermal radiation from the accretion disk and/or the Broad
Line Regions (External Compton, EC; Sikora, Begelman \& Rees 1994;
Dermer \& Schlickeiser 1993; Ghisellini \& Madau 1996).

However, irrespective of the origin of the seed photons, the
synchrotron plus IC interpretation implies that there must be close
correspondence between the synchrotron and inverse Compton portions of
the spectrum, since both are produced by the same electron
population. For a one-zone, homogeneous model where a population of
relativistic electrons, distributed as a broken power law with break
energy $\gamma_{\rm peak}$, emits synchrotron radiation with a peak at
$\nu_S$, the corresponding peak of the IC flux must be at $\nu_{SSC}
\propto \gamma_{\rm peak}^2 \nu_S$ for the SSC process, or $\nu_{EC}
\propto \gamma_{\rm peak}^2 \Gamma \nu_0$ for the EC case (where
$\Gamma$ is the bulk Lorentz factor, and $\nu_0$ the frequency where
the external radiation peaks).  Moreover, variability of the
synchrotron flux should be accompanied by variability of the
corresponding IC flux; the amplitude of the latter is different in the
various models (e.g., Ulrich et al. 1997) and can be used in principle
to put constraints on the origin of the seed photons.  Inspection of
Figure 5 shows that, at least in the case of 2200+420 where the data
coverage is better, the SED exhibits two broad peaks, one at $\sim
10^{14}$ Hz (\S~3.4), which we identify with the synchrotron
component, and a second one between 100 MeV and a few GeV, which can
be attributed to the inverse Compton component.
 
From the data in Figure 5, it is possible to derive an order of
magnitude for the magnetic field, $B$ (in Gauss), and the electron
energy $\gamma_{\rm peak}$, for both the SSC and EC models (where
$\delta$ is the beaming factor):

\begin{equation} 
(\gamma_{\rm peak}^{SSC})^2 = {3\nu_C \over 4\nu_S}
\end{equation}

\begin{equation} 
B^{SSC} = \frac{1+z}{\delta} \frac{\nu_S^2}{2.8 \times 10^6 \nu_C} 
\end{equation}

\begin{equation} 
(\gamma_{\rm peak}^{EC})^2 = {3 \nu_C \over 4 \nu_0}{1+z \over \delta\Gamma}
\end{equation}

\begin{equation} 
B^{EC} = {\Gamma \nu_S \nu_0 \over 2.8\times 10^6 \nu_C}
\end{equation} 

\noindent 
(e.g., Sambruna et al. 1997 and references therein). As an example we
now apply these relations to the 1997 SED of 2200+420, which is the
best constrained.  If we set, from Figure 5, $\nu_S=10^{14}$ Hz and
$\nu_C=10^{24}$ Hz, we derive $\gamma_{\rm peak}^{SSC}\sim 9\times
10^4$ and $B^{SSC}=3.8\times 10^{-3}/\delta$ Gauss.  Within the SSC
scenario, the ratio $L_C/L_S$ between the $\gamma$-ray and the optical
synchrotron luminosity is equal to the ratio between the synchrotron
and the magnetic energy density.  Setting $L_C/L_S\sim 3$ from Figure
5 and measuring the size through the variability time-scale, $R\sim
ct_{var}\delta$, in the SSC case we obtain:

\begin{equation} 
t_{var} \sim \left( {2 L_S \over 3 c^3 \delta^6 B^2}\right)^{1/2}
\sim 4.8\times 10^3 L_{S,44} \delta^{-2}\,\,\,\, {\rm days}
\end{equation} 

\noindent
where $L_{S,44}$ is the observed synchrotron luminosity in units of
$10^{44}$ erg s$^{-1}$.  The above relation shows that the SSC model
has difficulties in explaining spectra where the ratio $\nu_c/\nu_s$
is large (like in the 1997 July SED of 2200+420), since in this case
the resulting small magnetic field implies very large regions and/or
implausibly large values of the Doppler factor: a variability
timescale of $\sim$1 day implies, from Eq. (5), $\delta\sim 70$.

The recent detection of a broad and variable H$\alpha$ emission line
in 2200+420 (see \S~1) may provide an argument in favor of the EC
model.  Application of Eqs. (3) and (4) for the EC model (assuming
$\nu_0=10^{15}$ Hz) gives a larger magnetic field, $B^{EC} = 0.036~
\Gamma$ Gauss, than the SSC model, implying the electrons at the
synchrotron peak are less energetic, $\gamma_{\rm peak}^{EC}
=2.7\times 10^4/(\Gamma\delta)^{1/2}$.  In this case the ratio
$L_C/L_S$ is controlled by the ratio of the external radiation energy
density, $U_{ext}^\prime$ (as seen in the synchrotron-emitting blob
comoving frame), and the magnetic energy density.  If the external
radiation is produced by the broad lines, then $U_{ext}^\prime \sim
L_{BLR}\Gamma^2/(4\pi R_{BLR}^2c)$, where $L_{BLR}$ is the luminosity
of the broad line region of size $R_{BLR}$.  To produce $L_c/L_S\sim
3$, we must have

\begin{equation} 
L_{BLR} \sim { 3 R_{BLR}^2 B^2 c \over 2\Gamma^2} 
\end{equation} 

\noindent
which gives $L_{BLR} \sim 1.3\times 10^{41}$ erg s$^{-1}$ for
$R_{BLR}=5\times 10^{16}$ cm (e.g., Dietrich et al. 1998; Ulrich et
al. 1997), and for the same values as above for $\nu_0$, $\nu_c$ and
$\nu_S$.  This value is close to the emission line luminosity seen in
May and June 1995 (Vermeulen et al. 1995).  Note also that there may
be other contributions to the external radiation field, coming by the
accretion disk, by the intra-cloud material that scatters radiation
coming from the accretion disk, and by a small portion of the broad
line region illuminated by the jet.

For the 1995 SED, assuming $\nu_S=10^{14}$ Hz, $\nu_C=10^{22}$ Hz (as
indicated by the non-simultaneous EGRET spectrum), 
we obtain higher magnetic fields and lower electron energies for both
SSC and EC.

\subsubsection{Fits to the SEDs with Homogeneous Models} 

We fitted the SEDs of 1749+096 and 2200+420 with a one-zone
homogeneous model, assuming that the radio through UV emission is
synchrotron radiation from a relativistic electron population and that
the high-energy continuum is produced by upscattering off the same
electrons of photons either internal (SSC) or external (EC) to the
jet. Since the IR to UV/X-ray synchrotron emission is optically thin,
these energies could be produced in a single homogeneous region. An
additional spatial component, identified with different jet regions
with different self-absorption turnovers, is assumed to emit the radio
through submm portion of the spectra.

The external radiation is assumed to be narrowly distributed around a
typical frequency $\nu_0 \approx 10^{15}$ Hz.  The emission region is
uniformly filled with a plasma of electrons of random energy $\gamma$
and constant bulk motion with Lorentz factor $\Gamma$ in a constant
magnetic field $B$.  The electrons are continuously injected at a rate
$Q(\gamma)\propto \gamma^{-s}$ between minimum and maximum energies
$\gamma_{min}$ and $\gamma_{max}$.  The steady particle distribution
is computed self-consistently considering radiative cooling (including
the effects of the Klein-Nishina cross-section) and electron-positron
pair production. The latter effect is however unimportant given the
very small intrinsic compacteness (the luminosity to size ratio).

Table 4 lists the parameters assumed to reproduce the 1995 quiescent
state and the 1997 flare state of 2200+420, and the 1995 SED of
1749+096, including the parameters of the radio-to-submm emitting
region (labelled ``Large'' in Table 4). The solutions are plotted on
the data in Figures 4 and 5. Our contemporaneous 1995 SEDs are
underconstrained at the higher energies, where no measurements at GeV
energies are available. For 2200+420, the non-contemporaneous 1995
January EGRET spectrum was used to set an upper limit to the power at
GeV energies during 1995 November.

\vskip 0.5 true cm
{\it 1749+096}
\vskip 0.5 true cm

The SED of 1749+096 can be fitted by the SSC model without any
contribution from external radiation.  However, the EGRET upper limit
does not constrain the high energy part of the SED well enough to draw
any strong conclusion.  Note that the synchrotron emission peaks at
relatively low frequencies, implying small values of $\gamma_{\rm
peak}$.  This in turn implies that the second order Compton process is
not completely inhibited by Klein-Nishina effects, and can dominate
the emission at the highest frequencies (see Fig. 4), while the
spectrum in the ASCA band corresponds, in the fit shown in Figure 4,
to the peak of the first order Compton process.

Because of the spectral correspondence implied by the synchrotron plus
IC model, in the SED in Figure 4 the soft X-rays are produced by the
same electrons emitting the radio-submm energies. The IC emission of
these electrons, as can be seen by the SSC fit in Figure 4, may not be
negligible in the soft X-ray band during low states of the source,
diluting any fast variability at these energies, consistent with the
absence of flux variability in our ASCA SIS data. 

\vskip 0.5 true cm
{\it 2200+420}
\vskip 0.25 true cm

For 2200+420, the lower state of 1995 is obtained assuming that an
intrinsic power of the relativistic electrons $L^\prime_{\rm
inj}=1.3\times 10^{41}$ erg s$^{-1}$ is injected within a region of
size $R=7\times 10^{15}$ cm. This corresponds to an intrinsic
compactness $\ell_{\rm inj}=6.7\times 10^{-4}$, to be compared to the
compactness of external radiation, as seen in the comoving frame of
the blob, of $\ell_{\rm ext}=2\times 10^{-5}$.  In this case the
locally produced synchrotron radiation dominates the Compton cooling,
and all but the highest frequencies are produced by the SSC process.

In contrast, the flaring state of the summer of 1997 can be fitted
assuming that the external radiation field increased by two orders of
magnitudes, providing the main contributor of seed photons at
relatively high (UV) energies.  As a consequence, the EC process
dominates at energies larger than a few MeV, producing a flat
spectrum. On the other hands, the SSC mechanism is still
predominant in the X-ray range (see Fig. 5).  The intrinsic injected
power, in this case, has been increased by a factor 15 with respect to
the value of 1995.

Note that, in principle, one can increase the ratio between the
external and the synchrotron radiation by increasing $\Gamma$, since
$U^\prime_{\rm ext}\propto \Gamma^2$, while the synchrotron radiation
energy density $U^\prime_{\rm syn}\propto L_{\rm syn}\delta^{-4}$.
Assuming a constant broad line emission, the 100-fold increase in
$U^\prime_{\rm ext}$ can be achieved by a 10-fold increase in
$\Gamma$, which implies an enormous change in the bulk properties of
the jet.

We conclude that the flaring state of 1997 has been produced by the
increase of both the amount of the external radiation and the power
injected in the jet.  Together with the observational evidence of
varying emission line luminosity in 2200+420, this implies that the
power channeled in the jet and the ionizing radiation are linked
together.  If any synchrotron (and inverse Compton) flare is linked to
an increase of the ionizing luminosity coming from the accretion disk,
then the enhanced emission lines are best observable soon after the
end of the non-thermal flare, since the time-scale of the broad line
variability is of the order of $R_{\rm BLR}/c\sim$ 1 month, which can
be significantly longer than the non-thermal flare. This prediction is
easily testable observationally even in the absence of GeV data, by
monitoring both the line and non-thermal continuum luminosity. 

There is tentative evidence for a possible time lag between the 1997
July optical and GeV flares (Bloom et al. 1998), with the shorter
wavelengths leading the longer ones by a few hours. Such a lag, if
true, would represent a difficulty for both SSC and EC models, since
in both cases one would naively expect that the synchrotron flare
should anticipate the inverse Compton activity (unless some
fine-tuning of the parameters is invoked in both cases). However, the
sparse sampling of the data prevents any firm statement about the lag.
Correlated flux variability is potentially very powerful in putting
constraints to the current emission models; more intense and regular
monitoring of 2200+420 at both optical and GeV energies is needed
before reaching any conclusions.

\section{Conclusions} 

Our ASCA observations of two LBLs demonstrate the importance of the
X-ray band to test current unification models for blazars. This is
because the synchrotron and inverse Compton components, whose balance
regulates the HBLs/LBLs/FSRQs transition, cross in the X-rays;
observations in this spectral region can thus directly quantify the
relative importance of the two emission mechanisms in different blazar
classes. Indeed, together with published data for more blazars, our
ASCA observations support the current understanding that HBLs are
synchrotron-dominated, while LBLs and FSRQs are Compton-dominated.
Broad-band X-ray observations of larger and possibly complete samples
are needed to put stronger constraints. In particular, we need to
explore the X-ray and multifrequency continua of low-redshift, non-GeV
FSRQs which are currently underrepresented in the ASCA and SAX
archives.

Multifrequency spectral and flux variability is an essential test to
probe detailed emission mechanisms for the higher energies, and assess
the central question of the origin of the seed photons for the inverse
Compton scattering. Our results supports current claims that External
Compton scattering could be the dominant cooling mechanisms for
blazars with strong emission lines. Future monitorings of correlated
line and synchrotron continuum variability have the potential to put
interesting constraints on these models. In particular, an accurate
estimate of the size of the Broad Line Regions in blazars is seeked,
since this is a fundamental parameter to calculate the radiation
densities of the gas in which the relativistic jets are embedded.

\acknowledgements

RMS was funded through NASA contract NAS--38252. EJH acknowledges
financial support from NASA grant NGT-51152, a NASA Graduate Student
Researchers Program Fellowship at the University of Arizona, and
NAGW-4266, plus observing help from John and Marcia Hooper. RIK
acknowledges support by Fermi National Accelerator Laboratory under US
Department of Energy contract No. DE-ACO2-76CH03000. This work was
supported in part by NASA grant NAG5-2538.

\section{Appendix: Parabolic fits to the radio-to-X-ray data}

This section illustrates the parabolic fits to the radio-to-X-ray
spectral energy distributions of 1749+096 and 2200+420 referred to in
\S~3.4. The data in the $\log(\nu L_{\nu}$) vs. $\log\nu$ space were
fitted with a function of the type

\begin{equation}
y=ax^2 + bx + c
\end{equation}

\noindent following Sambruna et al. (1996). In the case of 2200+420,
the fit included the softer X-rays ($0.6-2$ keV) since we have
evidence that the latter are the high-energy tail of the IR/optical
synchrotron emission; on the other hands, in 1749+096 the X-rays are a
different emission component and were not included in the parabolic
fit.

The fitted parameters are: for 1749+096, $a=-0.2476 \pm 0.0005$,
$b=6.5070 \pm 0.0046$, and $c=3.5803 \pm 0.0142$, and for 2200+420,
$a=-0.0979 \pm 0.0002$, $b=2.8076 \pm 0.0029$, and $c=23.3717 \pm
0.0416$. The fitted curves are shown superposed to the data in Figure
A1. The derived peak frequencies $\nu_p$, peak luminosity L$_p$, and
total integrated luminosity L$_{sync}$ reported in \S~3.4 have
uncertainties 10\% (on $\nu_p$) and 35\% or less (on the
luminosities); the errors were obtained by calculating the ranges of
values allowed by the errors on the parabola parameters $a, b$, and
$c$.

Formally, the reduced $\chi^2$s of the fits are high ($>$ 2) and not
acceptable; they are mostly contributed to by a few datapoints in the
IR/optical regimes in both cases, which have small errorbars. More
complex spectral functions are probably needed to describe the data
more closely; however, given the sparse sampling (especially in the
region where the peak frequency falls, $\approx 10^{13-14}$ Hz), the
parabolic function provides the simplest and yet relatively accurate
approximation, useful to derive an order-of-magnitude estimate on
$\nu_p$, L$_p$, and L$_{sync}$.

\newpage

\noindent{\bf Figure Captions}

\begin{itemize}

\item\noindent Figure 1: ASCA spectrum of 1749+096 in 1995 September.
{\it (a):} Data and folded model, a single power law with photon index
$\Gamma=1.8$ plus Galactic column density; {\it (b):} Residuals,
plotted as the ratio of the data to the model. Data from all the four
detectors, which were fitted together, are plotted in the Figure. The
model provides an acceptable description of the ASCA data.

\item\noindent Figure 2: ASCA spectrum of 2200+420 in 1995 November.
{\it (a):} Data and folded model, consisting of a single power law
plus total (atomic + molecular) Galactic column density; {\it (b):}
Residuals, plotted as the ratio of the data to the model. Data from
all the four detectors, which were fitted together, are plotted in the
Figure. The model clearly underestimates the flux at the lower
energies (see also Figure 3). 

\item\noindent Figure 3: Confidence contours for the column density
N$_H$ and the photon index from the fits to the ASCA spectra of
1749+096 and 2200+420. The 68\%, 90\%, and 99\% confidence levels for
two interesting parameters are shown. The solid lines mark the
Galactic N$_H$ in the direction to the sources (total atomic +
molecular column for 2200+420); the dashed lines for 1749+096 are the
formal 90\% uncertainties. No range for the total Galactic column is
reported for 2200+420, since the uncertainties on the molecular
contribution are not known. For 1749+096, excess column density over
Galactic is present at $\sim$ 98\% confidence, while in the case of
2200+420 the fitted column density is lower than Galactic, providing
evidence for soft excess in this source.

\item\noindent Figure 4: Multifrequency spectral energy distributions
of 1749+096.  The filled symbols are the ASCA data and the
contemporaneous radio and optical observations obtained by us in 1995
September, while the open symbols are non-simultaneous literature data
(see Sambruna et al. 1996 for references), including two ROSAT
observations (Urry et al. 1996). The source was not detected at GeV
energies; the arrow marks the EGRET sensitivity threshold. Plotted as
a solid line is the best-fit to the 1995 data with a homogeneous,
one-zone synchrotron-self Compton model; the dashed line is the
contribution to the radio spectrum from a more extended region
(parameters are reported in Table 4).

\item\noindent Figure 5: Multifrequency spectral energy distributions
of 2200+420.  The filled symbols are our ASCA data and contemporaneous
radio and optical observations in 1995 November, together with the TeV
upper limit measured in the same period (Catanese et al. 1997).  The
open circles are the EGRET spectrum measured in 1995 January (Catanese
et al. 1997).  Plotted with triangles and open ``bow-ties'' in X-rays
and $\gamma$-rays are the simultaneous data during the optical/GeV
flare in 1997 July (Bloom et al. 1998; Madejski et al. 1997; Makino et
al. 1997). The squares are the data from the 1988 multifrequency
campaign (Bregman et al. 1990; Kawaii et al. 1991). The solid line is
the best-fit to the 1995 data with a homogeneous, one-zone
synchrotron-self Compton model, while the dotted line is the
contribution to the radio spectrum from a more extended region
(parameters are reported in Table 4). During the ``quiescent'' state
of 1995 November, the SSC process is primarily responsible for the
production of the flux from radio to X-rays. In contrast, the flaring
state of 1997 July requires an increase of the external radiation
density and the power injected in the jet (Table 4), resulting in the
dominance of the EC process at energies larger than a few MeV, while
the SSC process still dominates the power output at X-rays and longer
wavelengths. This model is plotted with a dashed line in the Figure.

\item\noindent Figure A1: Contemporaneous spectral energy
distributions from radio to X-rays of 1749+096 and 2200+420 in
1995. The dotted lines are the fits with a simple parabolic function
(see Appendix). For 2200+420 the soft X-rays ($0.6-2$ keV) were
included in the fit, as the spectral slope in this energy range is
consistent with the extrapolation from the optical.  For 1749+096 the
X-rays were excluded from the parabolic fit since they originate
through a different emission mechanism than the lower frequencies. The
synchrotron peak frequency in both sources falls in a poorly sampled
spectral region ($\approx 10^{13-14}$ Hz). The fitted parabola
parameters (see Appendix) were used to calculate the peak frequencies,
peak luminosities, and the total integrated synchrotron luminosity
(\S~3.4).

\end{itemize}

\end{document}